# Defending the Quantum Reconstruction Program

Philipp Berghofer

University of Graz

philipp.berghofer@uni-graz.at

**Abstract**

The program of reconstructing quantum theory based on information-theoretic principles enjoys much popularity in the foundations of physics. Surprisingly, this endeavor has only received very little attention in philosophy. Here I argue that this should change. This is because, on the one hand, reconstructions can help us to better understand quantum mechanics, and, on the other hand, reconstructions are themselves in need of interpretation. My overall objective, thus, is to motivate the reconstruction program and to show why philosophers should care. My specific aims are threefold. (i) Clarify the relationship between reconstructing and interpreting quantum mechanics, (ii) show how the informational reconstruction of quantum theory puts pressure on standard realist interpretations, (iii) defend the quantum reconstruction program against possible objections.

## 1. Introduction

The 21st century, so far, has been a blooming era for the field of quantum foundations. A growing number of physicists don't want to restrict their research to the task of calculation any longer, instead embracing a "shut up and contemplate!" attitude (Hardy and Spekkens 2010). This is to say that some physicists have come to appreciate the value of philosophy – particularly when it comes to rigorous conceptual analysis and foundational thinking. This has manifested in many fruitful interdisciplinary collaborations and much progress has been achieved at the intersection of physics and philosophy. This is not to deny that the infamous shut up and calculate attitude is still dominant in most physics departments. This is not even to say that the shut up and calculate attitude does not often help to make progress in physics. This is stating the simple fact that physicists and philosophers have much to say to each other and that the field of quantum foundations has significantly benefited from this dialogue.

Here is a further fact: In the course of the 21st century, many physicists working on quantum foundations have shifted their attention away from the project of *interpreting* quantum theory to the project of *reconstructing* it. This has been referred to as "the new wave of quantum foundations" (Chiribella & Spekkens 2016) as well as a "paradigmatic shift in the domain of the foundations of physics" (Grinbaum 2006, 2007). And here is another fact: Philosophers have not cared much about this development. This is startling. It is startling for several reasons. First, the project is not completely new but emerged at the turn of the millennium as a consequence of the booming interest in quantum information. Second, the researchers involved in this project are physicists such as Carlo Rovelli and Anton Zeilinger who are well-known for significantly contributing to philosophy of physics. Third, its basic idea is simple and prima facie plausible, easily backed up by prominent historical examples. Fourth, the promise of successful reconstructions is that they have crucial implications for understanding quantum mechanics. Fifth, even if this promise can be fulfilled, reconstruction cannot be the end of the story. Reconstructions, as I will argue, are themselves in need of interpretation. The first two points imply that philosophers of physics know about this project. The last three points imply that philosophers should care about this project. This raises the question of why this program has received so little attention in the philosophy community. This is a question I do not attempt to answer in this paper. I assume that it does not concern so much the fact that reconstruction is different from interpretation but more that the existing successful reconstructions seem to put pressure on the most prominent interpretations we find in philosophy of physics.

So what is the objective of my paper? My objective is threefold. 1. Clarifying the relationship between reconstructing and interpreting quantum mechanics. 2. Showing how the project of reconstructing quantum theory based on information-theoretic principles puts pressure on so-called "quantum theories without observers" which are the most prominent interpretations in the philosophy community. 3. Defending the quantum reconstruction program. The paper is structured as follows. Section 2 introduces and motivates the quantum reconstruction program. In Section 3.1, I clarify my view on the relationship between reconstruction and interpretation. Section 3.2 argues that reconstructions are themselves in need of interpretation. Section 3.3 shows why successful information-theoretic reconstructions are in tension with the standard interpretations we find in philosophy and instead support interpretations that deny that the wave function represents a physical state. In Section 4, I defend the quantum reconstruction program against possible objections.

## 2. Motivating the quantum reconstruction program

The cornerstones of the quantum reconstruction program (QRP) have been independently formulated by Carlo Rovelli (1996) and Anton Zeilinger (1999). In both works, it is explicitly argued that quantum mechanics needs to be based on a set of simple *physical* principles. Both suggest concrete *information-theoretic* principles that could play such a role.[1] And both mention *special relativity* as a role model in this regard: a physical theory that has counter-intuitive consequences but is widely accepted since it conceptually rests on clear physical principles. Particularly Rovelli's paper captures precisely the spirit, approach, and ambition of QRP:

> Quantum mechanics will cease to look puzzling only when we will be able to *derive* the formalism of the theory from a set of simple physical assertions ('postulates,' 'principles') about the world. Therefore, we should not try to *append* a reasonable interpretation to the quantum *mechanics formalism,* but rather to *derive* the formalism from a set of experimentally motivated postulates. (Rovelli 1996, 1639)

The success of and booming interest in quantum information theory convinced more and more researchers that the notion of information is crucial for understanding the foundations of quantum mechanics. In the year 2000, Christopher Fuchs and Gilles Brassard co-organized a conference with the programmatic title "Quantum Foundations in the Light of Quantum Information." Fuchs' paper of the same name has been highly influential, summarizing the methodology of this project as follows: "to reduce quantum theory to two or three statements of crisp *physical* (rather than abstract, axiomatic) significance. In this regard, no tool appears to be better calibrated for a direct assault than quantum information theory" (Fuchs 2001). Soon after the conference, the first successful reconstruction was offered by Hardy 2001. Several others followed, e.g., Chiribella et al. 2011, Dakic & Brukner 2011, Masanes et al. 2013, Goyal 2014, Höhn 2017, and Höhn & Wever 2017, and the reconstruction program continues to shape the field of quantum foundations (as exemplified by Chiribella & Spekkens 2016a and D'Ariano et al. 2017). As mentioned above, it is certainly a virtue of this program that it is based on a simple and convincing idea. In what follows in this section, I shed some light on this idea and its main motivation.[2]

---

[1] Rovelli identifies two principles (1996, 1657f.), Zeilinger one (1999, 635).

The idea is that instead of taking the quantum formalism as a given and trying to make sense of it by contemplating the ontological status of the mathematical terms involved, such as, most prominently, the wave function, we should be looking for foundational physical principles from which the formalism[3] can be derived or reconstructed. The point is that the formalism of quantum mechanics is couched in highly technical terms that complicate a direct interpretation. More precisely, it is couched in terms of Hilbert spaces and self-adjoint operators such that the quantum state is represented by a vector in Hilbert space and observables are represented by self-adjoint operators. Now, the spirit of the reconstruction program is that instead of asking how to ontologically interpret the mathematics, we want to know where the mathematics comes from. Why is nature so successfully described by the mathematics of complex Hilbert spaces? The idea is that this question can only be answered by deriving or reconstructing the formalism from principles that have a clear physical meaning.

> In short, the postulates of quantum theory impose mathematical structures without providing any simple reason for this choice: the mathematics of Hilbert spaces is adopted as a magic blackbox that 'works well' at producing experimental predictions. However, in a satisfactory axiomatization of a physical theory the mathematical structures should emerge as a consequence of postulates that have a direct physical interpretation. By this we mean postulates referring, e.g., to primitive notions like physical system, measurement, or process, rather than notions like, e.g., Hilbert space, $C^*$-algebra, unit vector, or self-adjoint operator. (D'Ariano et al. 2017, 1)

We will discuss the relationship between reconstruction and interpretation in more detail in Section 3. However, compare the above-quoted sentiment with the approach most popular in philosophy of physics: "A precisely defined physical theory […] would never use terms like 'observation,' 'measurement,' 'system,' or 'apparatus' in its fundamental postulates. It would instead say precisely

---

2   Of course, this is only a very rough glimpse at the history of quantum reconstruction. In fact, operational axiomatizations can be traced back to von Neumann himself and his joint work with Garrett Birkhoff on quantum logic. It has been pointed out in this context that von Neumann confessed to Birkhoff that he did not "believe in Hilbert space anymore" (D'Ariano et al. 2017, 2; Grinbaum 2017). The quantum reconstruction program also significantly drew on developments in probability theory. The QBist reconstructions were significantly influenced by the work of Bruno de Finetti (see, e.g., Fuchs 2001). The reconstructive work offered in Goyal et al. 2010 was influenced by Cox' derivation of probability theory from Boolean algebra.

3   By "formalism" I understand the Hilbert space formalism of quantum theory as it has been established by von Neumann. More precisely, most informational reconstructions seek to derive the *finite*-dimensional Hilbert space formalism. Here is how Chiribella and Spekkens put it: "What distinguishes the axiomatic work being pursued today is the use of notions inspired by quantum information theory, the emphasis on composite systems, the focus on finite-dimensional Hilbert spaces, and an insistence on axioms that are operationally meaningful" (Chiribella & Spekkens 2016b, 4).

*what exists and how it behaves*" (Maudlin 2019, 5). This illustrates that taking the formalism as a given is how the project of interpreting quantum mechanics typically proceeds. The many-worlds interpretation argues that we can make sense of the quantum formalism straightforwardly by taking it at face value and assuming that the wave function always obeys the Schrödinger equation. Bohmian mechanics argues that we can make sense of the formalism only by modifying it first and then taking the new formalism at face value, reading off a particle ontology. By contrast, the underlying rationale of the reconstruction program is that we can gain a better understanding of quantum theory by first identifying foundational principles from which the formalism can be derived.

The most widely discussed reconstructions quoted above essentially employ the following two-step procedure: Introduce a framework that covers all possible probability theories and specify a few simple physical principles for which it is shown that the quantum formalism uniquely follows from these principles (see Koberinski & Müller 2018, 260). Why does quantum mechanics have the mathematical structure it has? Because it is the only possible mathematical structure that satisfies the respective meaningful principles. These principles typically are information-theoretic principles. Accordingly, the program promises to fulfill two of John Archibald Wheeler's main ideas. First, it seeks to answer Wheeler's famous question "Why the quantum?" Second, it is in line with Wheeler's infamous postulate "it from bit."

I take it that all of this is prima facie plausible. Still, there are a number of worries one might raise. Section 3 intends to clarify the relationship between reconstruction and interpretation, which will be helpful for getting a better understanding of QRP. In Section 4, I discuss several possible objections against QRP. First, however, let us address what its proponents typically view as a prime example of a successful reconstruction as well as a main motivation for the *quantum* reconstruction program (see, e.g., Rovelli 1996, 1639; Zeilinger, 1999; Chiribella & Spekkens 2016, 3). This is Einstein's principle-based approach to special relativity.

In this context, two curious features of special relativity are particularly important to us. The first one concerns the fact that special relativity has highly counter-intuitive implications but is widely accepted among physicists and philosophers. We don't observe Lorentz contractions or time dilatations in our daily lives. Most importantly, the relativity of simultaneity is in clear tension with our classical understanding of time. Still, contemporary philosophy of physics is not dominated by the topic of interpreting special relativity, as it is the case with quantum mechanics. We don't have fancy quotes about nobody understanding special relativity. Instead, Einstein's 1905 understanding of special relativity is widely accepted.[4] Why is this? Importantly, the answer is not that no other interpretation of its mathematical content is possible. Lorentz, for instance, believed in the aether long after 1905. And he did so in a way that is consistent with the mathematical structure of special relativity, a mathematical structure that was introduced by Lorentz himself several years before Einstein's 1905 paper. This brings us to the second curious feature of special relativity. Namely precisely the fact that the mathematical formalism of special relativity was already present in Lorentz' work several years before Einstein's annus mirabilis. Still, Einstein is generally considered the father of special relativity. What is it that makes his contribution special?

Accordingly, we have two questions. Why is it that the counter-intuitive consequences of special relativity are widely accepted while quantum mechanics is plagued by a plethora of conflicting interpretations?[5] What distinguishes Einstein's treatment of the mathematical formalism of special relativity from earlier works that introduced the same mathematics? The idea is that both questions have the same answer: In 1905 Einstein succeeded in deriving the mathematical formalism of special relativity from two meaningful physical principles: the light postulate and the principle of relativity. We (believe to) understand special relativity because we understand the underlying principles. We (or most of us) accept Einstein's "interpretation" of the mathematics encoded in the Lorentz transformations because the empirically well-tested formalism derives from these principles and we regard the principles more plausible than, for instance, more complicated aether theories.

Of course, this depiction of the difference between quantum mechanics and special relativity is a huge oversimplification. I return to this topic in a bit more detail in Section 4.1. However, the above suffices in clarifying that the reconstruction program is based on a plausible, almost trivial, idea: It is more straightforward to understand the physical meaning of physically meaningful principles than to understand the physical meaning of an abstract mathematical formalism. Still, successful reconstructions (as well as Einstein's two postulates) should not be viewed as endpoints of our understanding but rather as starting points in our exploration.[6] The next section is intended to shed more light on the relationship between reconstruction and interpretation.

---

4  Of course, this is not to say that subsequent work on special relativity did not lead to important advances in our understanding of the theory (e.g., the work of Minkowski). And I also do not want to suggest that there are no important approaches that question the orthodox understanding of special relativity (see, e.g., Brown 2005, Brown & Read 2022, Knox 2018). Crucially, I believe that any scientific theory is in need of interpretation which will become important in Section 3.2.

5  One may argue that this is a false equivalence, claiming that we do accept the counter-intuitive consequences of quantum mechanics but simply lack a physical interpretation of them. However, this would be to downplay how severely different interpretations contradict each other. The most prominent interpretations in philosophy, Bohmian mechanics, the many-worlds interpretation, and GRW come to very different conclusions about *which* counter-intuitive consequences we should accept. Indeterminism is implied by GRW (and other objective collapse theories) but avoided in Bohmian mechanics (although this has recently been challenged by Landsman 2022) and MWI. Non-locality is implied by Bohmian mechanics and GRW but avoided in MWI. Also, all three of these interpretations explicitly reject one or more key postulates of textbook quantum mechanics. Bohmian mechanics and GRW actually modify the formalism of quantum mechanics. MWI denies the collapse postulate. This is very different from the situation in special relativity. For instance, no mainstream interpretation of special relativity would contradict the light postulate.

## 3. On the relationship between reconstruction and interpretation

### 3.1. Basic ideas and options

As pointed out above, according to QRP we should not take the quantum formalism as a given, attempting to read off the nature of physical reality from the mathematical structure of the formalism. This is how proponents of QRP typically view the project of interpretation. They tend to be particularly critical regarding currently popular versions of wave function realism, arguing that in contemporary philosophy of quantum mechanics "the strategy has been to reify or objectify all the mathematical symbols of the theory and then explore whatever comes of the move" (Fuchs & Stacey 2019, 136). Instead, QRP is a project of "taking a more physical and less mathematical approach" (Masanes et al., 2013, 16373). More precisely, QRP is the program of deriving the quantum formalism from a set of operationally meaningful principles. The underlying idea is that only by means of reconstruction we can hope to adequately understand quantum theory. This brings us to the following question: What precisely is the relationship between the quantum reconstruction program (QRP) and common interpretations of quantum mechanics, i.e., the project of quantum interpretation (PQI)? Here are some options:

> O1: QRP is simply a novel interpretation among interpretations such as the many-worlds interpretation and Bohmian mechanics.
>
> O2: QRP is *inconsistent* with PQI in the strong sense that if one program is successful, the other one is not. If QRP succeeds, any existing interpretation must be considered wrong.
>
> O3: QRP is *in strong tension* with PQI in the following sense: If QRP is successful, then PQI is superfluous, there is no need for (further) interpretations.
>
> O4: QRP and PQI are more or less *unrelated* projects. QRP is equally consistent with any existing interpretation.
>
> O5: QRP and PQI are *complementary* projects in the following sense: 1. The success of one does not rule out the success of the other. 2. Both projects can benefit from each other. 3.

---

6   I'm grateful to Philip Goyal for putting this in these helpful terms in personal conversations.

Reconstructions are themselves in need of interpretation. 4. QRP has some strong implications for PQI. 5. In particular, QRP puts pressure on objectivist[7] and $\psi$-ontic[8] interpretations.

My main objective in what follows is to motivate and defend O5. Before I focus on O5, I quickly comment on the other options.

O1 is wrong for reasons that should be obvious by now. Interpreting the quantum formalism and reconstructing it are different projects.[9] Of course, one may argue that QRP leads to a distinctive information-based interpretation. The falsehood of O2 is also rather obvious. Reconstructing the formalism is not inconsistent with also interpreting it. However, proponents of QRP might feel sympathetic to O3 (see Grinbaum 2007). Here the idea is that the ultimate goal is to find a reconstruction in terms of physical principles that are so clear that no further interpretation is necessary. When I discuss O5 in more detail, it will become clear why I find O3 implausible. O4 seems to be a safe choice. In this context, it is useful to refer to the distinction Chiribella and Spekkens make between the dynamicist and the pragmatist tradition in physics:

> Within the dynamicist tradition, the physicist's job is to describe the natural dynamical behaviour of a system, without reference to human agents or their purposes. In the pragmatic approach, on the other hand, the laws of physics are characterized in terms of the extent to which we can learn and control the behaviour of physical systems. (Chiribella & Spekkens 2016, 2)

Chiribella and Spekkens seem to conceive of these approaches as rival perhaps mutually exclusive options, pointing out that "[t]he new foundational work [of QRP] represents a renewed interest in exploring quantum theory within the pragmatic tradition" (Chiribella & Spekkens 2016, 2). But why not believe that both approaches are perfectly consistent? In the spirit of O4 one may argue that interpretation (dynamicist approach) tells you how the world works and the informational reconstructions (pragmatist approach) are useful for the experimental physicist. Perhaps this is an implicit attitude philosophers of physics have toward QRP. However, proponents of QRP would typically insist that QRP delivers crucial insights into the nature of quantum theory (and, arguably, the world that is so successfully described by it). Stressing the analogy to special relativity again: If you believe that the relativity principle is a basic physical principle that can be used to derive special relativity, you would not say that special relativity is consistent with the view that there is a privileged frame of reference. Similarly, proponents of QRP "characterize quantum theory *as a theory of information*" (D'Ariano et al. 2017, 5). The lesson that for instance Jeffrey Bub draws from this is that we should interpret "quantum theory as a theory about the representation and manipulation of information, which then becomes the appropriate aim of physics, rather than a theory about the ways in which nonclassical waves or particles move" (Bub 2004, 243). The way I see it, the more evidence we have that quantum mechanics is, fundamentally, about information, the less reason we have to believe in a dynamicist interpretation that seeks to purge quantum mechanics of all operational notions. This is why, in my terminology, I argue that the success of QRP puts pressure on objectivist (~dynamicist) interpretations. We return to this in subsection 3.3.

---

7  By objectivist interpretations I understand interpretations that claim to deliver a purely objective third-person perspective that does not contain any irreducibly subjective/operational concepts or perspectival moments.
8  Here and in what follows, I use the term "$\psi$-ontic interpretation" in the informal sense of covering any interpretation according to which the wave function represents an ontically real state. This is to be distinguished from the technical sense in which this notion has been introduced by Harrigan and Spekkens (2010). The three main objectivist interpretations, MWI, BM, and GRW, are also $\psi$-ontic interpretations in this informal sense.
9  Unfortunately, the notion of "interpretation" is ambiguous and not consistently used in the literature. For the present purposes, it suffices to keep in mind a clarification suggested by an anonymous reviewer of this journal: Reconstructions and interpretations "are not at the same level of abstraction. While QRP aims at understanding QT by reconstructing it, a (genuine) interpretation takes the formalism for granted, and is more 'metaphysical': it aims at giving an ontology, understanding Nature."

As mentioned, option O5 is the one I defend. I take it that the claims 3 and 5 of O5 are particularly controversial. In the next subsection, I defend claim 3, arguing that reconstructions are themselves in need of interpretation.

**3.2. Why reconstructions are in need of interpretation**

The first thing to note is that at the present moment there exist several different reconstructions of quantum theory. The common attitude among proponents of QRP is that each successful reconstruction is valuable and contributes to shedding light on particular features of quantum mechanics.[10] Accordingly, it is not to be expected that there is one privileged reconstruction/axiomatization. This is analogous to how special relativity can be derived from Einstein's two postulates, or, alternatively, from the single postulate of Minkowski spacetime. The mere fact that there are several reconstructions/axiomatizations brings up several interpretative issues (contrast, e.g., Maudlin 2012 with Brown 2005). It can be argued that Minkowski's geometrical reconstruction sheds new interpretative light on special relativity. Thus, not every interpretative issue is resolved or prevented by Einstein's two principles being "very simple and intuitively clear" (Zeilinger 1999, 632). This is why at the end of Section 2 I expressed my sympathy with the view that reconstructions/axiomatizations should not be viewed as endpoints but rather as starting points of our understanding of physical theories.

---

10 "Every axiomatization has its own benefits" (Müller & Masanes 2016, 140). In fact, it can be argued that even unsuccessful reconstructions can teach us important lessons about quantum theory (Koberinski & Müller 2018, 277). For a dissenting view according to which there should be "a unique correct axiomatisation," see Adlam 2022.

The fact that our physical theories cannot be understood as offering a unique mathematical representation of the physical world has become clearer and clearer. This is most clearly seen in philosophical reflections on dual theories (see, e.g., Castellani and Rickles 2017 and De Haro and Butterfield 2021). However, we do not need to invoke the celebrated AdS/CFT conjecture to make this point. Even in classical mechanics we have the situation that the same physics can be formulated in different mathematical frameworks. Newtonian, Lagrangian, and Hamiltonian mechanics constitute different mathematical frameworks for solving the same physical problems. Are Newton's famous principles sufficiently clear that any further interpretations are superfluous? Philosophical discussions in classical mechanics show that they are not (see North 2022 and Wilson 2013).[11] For instance, classical $N$-particle mechanics can be expressed as a theory in three-dimensional Euclidean space or a theory in $3N$-dimensional configuration space (see Wallace 2021b, 68). What is the real space we live in? The obvious answer seems to be that space is three-dimensional and one might think the only philosophical problem left here is the question as to why space is three-dimensional. However, prominent voices such as Wheeler, the spiritual father of QRP, would insist that mathematical space must be distinguished from physical space and that mathematical models and concepts can only *approximate* physical reality (Wheeler 1980a, 1980b). In this view, it would even be wrong to identify the physical space we live in with three-dimensional Euclidean space. There can never be a perfect fit between physical reality and mathematical representation.

Of course, here my aim is not to convince the reader to subscribe to Wheeler's strong claims. I only want to point out that it is a decisive moment of QRP (inherited from Wheeler) that we must not confuse mathematics with physical reality. The fact that there exist several distinct theoretical reconstructions of quantum theory reinforces this point. In sum, this leads to two reasons why reconstructions are in need of interpretation. First, the mere fact that there is no unique reconstruction leads to several interpretative issues just as it is the case in special relativity and classical mechanics. Second, physical theories, even if they can be reconstructed from clear physical principles, always involve mathematical notions and concepts that are in need of interpretation. Quantum mechanics is a highly technical theory. It will remain exactly as technical even after the formalism has been reconstructed. Reconstructions are supposed to *explain* where the mathematics comes from, *not* to reduce it. Accordingly, even after we have reconstructed the formalism, interpretative questions remain about how to interpret the mathematics. For instance, what is the ontological status of the wave function? What does it represent? I assume it would be naive to expect that such questions can be entirely settled by reconstructions. Regarding the relationship between reconstruction and interpretation, Goyal recently suggested the following "*two-step reconstruction-based strategy*":

> "*1. Reconstruct the quantum formalism.* First, reconstruct the quantum formalism, with the specific goal of distilling *the full physical content of the formalism into physical principles and assumptions that can be expressed in* natural language and that are amenable to philosophical reflection.
> 
> *2. Interpret the reconstruction.* Second, reflect on the principles and assumptions of the reconstruction, bringing to bear whatever philosophical traditions may be appropriate." (Goyal 2023, 340)

---

11  What is more, as discussed in more detail in Section 4.1, all the main objectivist interpretations of quantum mechanics are in need of further interpretation. Most prominently, this concerns the ontological status of the wave function. For instance, there is no agreement among Bohmians on whether the wave function is physically real, should be interpreted nomologically, etc. A more specific example would be the open question of how to interpret probabilities in MWI. I am grateful to an anonymous reviewer of this journal for prompting me to emphasize this point at the outset.

While Goyal says, reconstruct the formalism and interpret the principles, perhaps it would be more appropriate to say: Reconstruct the formalism, then interpret the formalism in light of the principles (Berghofer 2022, 340).

## 3.3. Why successful informational reconstructions put pressure on objectivist and *ψ-ontic* interpretations

We already mentioned some of the features of quantum mechanics that make it difficult to interpret. While in classical mechanics the ontology consists of point particles whose evolution in time is governed by deterministic differential equations, in quantum mechanics the state of the system is described by a vector in Hilbert space, i.e., the wave function, whose evolution in time is also governed by a deterministic differential equation, the Schrödinger equation. As noted, the nature of the wave function is an especially tricky and controversial topic when it comes to interpreting quantum mechanics. However, what makes quantum mechanics particularly interesting and distinct from classical mechanics is that its dynamics (according to textbook quantum mechanics) is not exclusively governed by the deterministic Schrödinger equation. As is well known, the wave function evolves according to the Schrödinger equation *unless a measurement occurs*. Upon measurement, the wave function is said to collapse, the quantum state ceases to be in a state of superposition, and consequently, we observe a definite value. For the working physicist, the wave function is a tool that allows us to calculate the probability that a quantum measurement will yield a certain result according to the Born rule. For the philosopher, this raises the question as to why it is that the wave function collapses upon measurement.[12] However, what is most relevant for the present investigation is the simple fact that in textbook quantum mechanics the concept of measurement is central and irreducible which renders the theory non-objectivist as it constitutively includes an operational element.

Of course, this doesn't align with the views of many philosophers of science and physics. In philosophy of quantum mechanics, the prevailing interpretations are so-called "quantum theories without observers" (Dürr & Lazarovici 2020, viii; Goldstein 1998). Here the dominant view is that the way quantum mechanics is taught and understood in physics textbooks is not only misleading but plainly unscientific. This is precisely due to the centrality of the notion of "measurement" in textbook quantum mechanics (see, e.g., Dürr & Lazarovici 2020, viii). We already quoted Maudlin saying: "A precisely defined physical theory […] would never use terms like 'observation,' 'measurement,' 'system,' or 'apparatus' in its fundamental postulates. It would instead say precisely *what exists and how it behaves*" (Maudlin 2019, 5).

Importantly, it has become evident that eliminating the concept of measurement from quantum mechanics is *very* difficult and comes at a cost. It is widely accepted that for quantum theories without observers there are two options: Either you accept the many-worlds interpretation or you modify the quantum formalism. Many regard the first option as unacceptable because it violates the principle of ontological parsimony as well as the idea that science should not postulate entities that are in principle unobservable. The problem with the second option is that the existing modificatory interpretations, most notably Bohmian mechanics, are less successful in their predictive power than standard quantum mechanics (Wallace 2022). This is because for Bohmian mechanics, in contrast to standard quantum mechanics, we don't have a relativistic extension (see Goldstein 2021, Section 1.4 and Kofler & Zeilinger 2010), which limits its applicability to a narrower range of phenomena. Consequently, opting for the second option and embracing a modificatory interpretation seems to contradict the essence of science in the following manner: it entails prioritizing our (classical) intuitions over our most successful scientific theory. Many scientists would subscribe to a principle like this: If on the one hand you have a highly successful scientific theory and on the other hand a modification of it that is less developed and less successful in its predictive power, you should go with the first one even if it has counter-intuitive consequences. The modificatory interpretations fly in the face of this principle. It is thus no surprise that while Bohmian mechanics is highly popular in philosophy, it remains largely ignored in physics. To be sure, I do not want to suggest that the above-mentioned problems mean that objectivist interpretations, i.e., quantum theories without observers, are no viable option. So far, I have only highlighted some challenges we encounter when interpreting quantum mechanics.

---

12   This is particularly puzzling if you consider the collapse a physical process. Why should nature care about whether a measurement is performed? Even opponents of subjective interpretations of the wave function concede that "[a]ny approach according to which the wave function is not something real, but represents a subjective information, explains the collapse at quantum measurement perfectly: it is just a process of updating the information the observer has" (Vaidman 2014, 17; see also Leifer 2014, Section 2.4).

Now, of course, our question is: What does the quantum reconstruction program add to this debate? It is to be noted that proponents of QRP are physicists that do not necessarily engage in questions about interpretation. After all, this would be the job of philosophers. One takeaway of this paper is that philosophers should step up and discuss what interpretative lessons we should draw from QRP in general as well as from specific successful reconstructions. However, when proponents of QRP address philosophical implications, they tend to stress affinities between QRP and Bohr's Copenhagen interpretation (Zeilinger 1999, Fuchs 2001, Chiribella & Spekkens 2016, 2, Goyal 2012). Goyal, one of the physicists working on reconstructions who is most vocal about philosophical implications of QRP, explicitly argued that what quantum mechanics teaches us is that science, at a fundamental level, is not supposed to provide "a description of *reality in itself* [but] a description of *reality as experienced by an agent*" (Goyal 2012, 584). In my view, this perfectly captures the Copenhagen spirit in Bohr's sense. Of course, there are several well-known problems with Bohr's approach which is why proponents of QRP are often attracted to other Copenhagen-like interpretations such as relational quantum mechanics (Höhn & Wever 2017), QBism (Appleby et al. 2017, DeBrota et al. 2020), or phenomenological approaches (Berghofer et al. 2020, Goyal 2023).

With this little background on the quantum measurement problem in place, we can now address how QRP puts pressure on objectivist and *ψ-ontic* interpretations. Remember that in the previous subsection we said that QRP suggests the following approach to interpreting quantum mechanics: Interpret the principles from which quantum theory can be reconstructed. (Or alternatively: Interpret the quantum formalism in light of the underlying principles.) Now, the principles in question are *information*-theoretic principles. They are typically formulated in terms of information and knowledge and specify how an observer can acquire information (see, e.g., Höhn 2017, 22-24). We see that this is in clear tension with objectivist interpretations. Objectivist interpretations are interpretations that claim to deliver a purely objective third-person perspective that does not contain any irreducibly subjective/operational concepts or perspectival moments. The "quantum theories without observers" that are most prominent in philosophy qualify as objectivist interpretations. The tension is obvious because QRP operates within an operational framework and the respective information-based axioms do contain operational terms. Now, one might believe that this is simply a terminological dispute about how to formulate the axioms. When Chiribella and Spekkens make the above-mentioned distinction between the dynamicist and pragmatic traditions, they exemplify this by quoting the following two formulations of the second law of thermodynamics (Chiribella & Spekkens 2016, 2):

> "It is impossible to devise a cyclically operating device, the sole effect of which is to absorb energy in the form of heat from a single thermal reservoir and to deliver an equivalent amount of work."
>
> "Heat can never pass from a colder to a warmer body without some other change, connected therewith, occurring at the same time."

The first formulation uses operative language, the second does not. The point is that here a "translation" from the pragmatic/operational language to the dynamicist is quite straightforward. This is not the case when it comes to quantum mechanics. Consider the following rule we find in Höhn's reconstruction (Höhn 2017, 22):

> "The observer $O$ can always get up to $N$ *new* independent bits of information about the system $S$. But whenever $O$ asks $S$ a new question, he experiences no net loss in his total information about $S$."

I take it that it is not clear how this postulate can be translated in a non-operational language. In this subsection, I started by pointing out that formulating quantum mechanics in a language that avoids operational terms such as "measurement" comes at a cost. And indeed, so far all successful reconstructions have been performed within an operational framework. Accordingly, interpreting these principles suggests an understanding of quantum mechanics in operational terms. This is how QRP puts pressure on objectivist interpretations. To make this point as precise as possible, given the above, I argue that the success of informational reconstructions puts pressure on objectivist interpretations in a fourfold way:

    First, as discussed, the relationship between reconstruction and interpretation can be summarized as follows: Reconstruct the formalism, then interpret the principles from which the formalism has been reconstructed. (Alternatively: Reconstruct the formalism, then interpret the formalism in light of these principles.) Successful reconstructions are formulated within an operational framework and based on information-theoretic principles. Accordingly, if we interpret quantum mechanics in terms of these principles, we interpret it in operational terms in a way that suggests an understanding of quantum mechanics as being centered around the concept of information. This does not square well with the dogma of objectivist interpretations that physical theories must never use operational terms but instead must "say precisely *what exists and how it*

*behaves*" (Maudlin 2019, 5). For instance, Höhn interprets the implications of his reconstruction as supporting the "partial interpretation" that "quantum theory is a law book governing an observer's acquisition of information about physical systems" (Höhn 2017, 3), and Höhn and Wever reinforce this by saying:

> [T]he successful reconstruction from this perspective underscores the sufficiency of taking a purely operational perspective, addressing only what an observer can say about the observed systems, in order to understand and derive the formalism of quantum theory. Ontic statements about a reality underlying the observer's interactions with the physical systems are unnecessary. (Höhn & Wever 2017).

Second, relatedly, the more successful our informational reconstructions turn out to be, the more evidence we have that quantum mechanics, fundamentally, is about information, the less reason there is to assume that the wave function represents an underlying ontic state. For instance, this is how QBists understand the wave function (quantum state) based on their reconstructions in terms of informationally complete measurements.

> A quantum state encodes a user's beliefs about the experience they will have as a result of taking an action on an external part of the world. Among several reasons that such a position is defensible is the fact that any quantum state, pure or mixed, is equivalent to a probability distribution over the outcomes of an informationally complete measurement. Accordingly, QBists say that a quantum state is conceptually no more than a probability distribution. (DeBrota & Stacey 2019).

This exemplifies the close connection between reconstruction and interpretation and how the success of QRP can be understood as motivating and supporting approaches that contradict objectivist or *ψ-ontic* interpretations. To be sure, I do not want to say that QRP implies that only *ψ-doxastic*[13] interpretations such as QBism or *ψ-epistemic* interpretations are viable options. However, we remember that at the beginning of this subsection I noted that even proponents of *ψ-ontic* interpretations such as Vaidman emphasize that the measurement problem can be easily avoided if you interpret the "collapse" of the wave function as an update of information. It is also acknowledged that the Born rule, prima facie, invites an operational understanding of the wave function (Wallace 2021b, 63). Accordingly, the success of informational reconstructions puts pressure on *ψ-ontic* interpretations as follows: If we can reconstruct the quantum formalism from information-theoretic principles, and if interpreting the wave function in terms of information avoids the measurement problem, why should we subscribe to an objectivist interpretation that leads to the problems discussed at the beginning of this subsection. Of course, this is not to deny that objectivist interpretations also have a number of virtues. The modest goal is to clarify how QRP puts pressure on them.

---

13  I explain this terminology in Section 4.8.

Third, as elaborated in more detail in Section 4.5, it seems that quantum theory in particular is well-suited to be reconstructed in terms of *information*-theoretic principles (as opposed to, e.g., classical mechanics). Furthermore, attempts to derive the quantum formalism from simpler physical principles have existed since the 1930s but it was only with the rise of quantum *information* theory that reconstructions began to have a real impact. Thus I propose that in their quest to make sense of quantum mechanics, philosophers should pay more attention to information-based interpretations.

Fourth, assuming that informational reconstructions are particularly helpful in elucidating the quantum formalism, the objectivist is in the following inconvenient position: Regarding their interpretation of quantum mechanics, they want to avoid all operational language, but if they wish to derive the formalism from simpler principles, they need to use information-theoretic principles. By contrast, if you subscribe to some non-objectivist information-based interpretation, you are in the favorable position to work within a unified conceptual framework that allows you to do both: reconstruct and interpret the formalism (see also footnote 12 below).

A more indirect line of reasoning of how QRP puts pressure on *ψ-ontic* interpretations has recently been suggested by Koberinski and Müller. Highlighting the explanatory power of QRP, they argue that "*reconstructions represent a challenge for existing 'ψ-ontic' interpretations of quantum theory by highlighting a relative deficiency of those interpretations in terms of their explanatory power*" (Koberinski & Müller 2018, 262). Here is the challenge in more detail:

> None of Bohmian mechanics, Everettian quantum theory, or collapse theories fill the explanatory role of a principle theory. For example, Everettian quantum theory does *not* start with a broad general framework of 'theories of many worlds,' put simple principles on top of that, and prove that quantum theory is the unique theory of many worlds that satisfies these principles. (Koberinski & Müller 2018, 265)

This is the promise of QRP that I called explaining where the mathematics comes from. I agree with Koberinski and Müller that this is a particular virtue of QRP. I also agree that this puts pressure on $\psi$-*ontic* interpretations.[14] However, Koberinski and Müller argue that while this puts pressure on $\psi$-*ontic* interpretations, it supports $\psi$-*epistemic* interpretations. This is the obvious move. But it is well-known that $\psi$-*epistemic* interpretations are challenged by the PBR theorem, often considered to be ruled out by this no-go theorem. Where does this leave us? Is the relationship between $\psi$-*epistemic* interpretations and QRP so strong that we are forced to say that if the former are ruled out, the latter is also no viable project? I address this worry in Section 4.6.

I conclude this section by further commenting on the relationship between reconstruction and interpretation. An anonymous reviewer of this journal pointed out that "explanation" is a multifaceted notion, and argued that there is no tension between successful information-based reconstructions and objectivist interpretations because the two projects ask different questions and pursue different explanatory strategies to make sense of quantum theory. Referring to footnote 12, in which I argue that Jessica is in a more favorable position than Peter, the reviewer says that there is no tension for Peter because "Peter is simply committed to the existence of two different ways of 'making sense' of the formalism." I agree that the two projects are concerned with different questions. In the context of interpretation, we typically encounter questions like these: What does the quantum formalism tell us about reality? In particular, what is the nature of the wave function?

---

14 An anonymous reviewer of this journal emphasized that $\psi$-*ontic* interpretations are not aimed at deriving the quantum formalism but at clarifying "what kind of reality could be associated with it." Accordingly, so the reviewer argues, proponents of $\psi$-*ontic* interpretations cannot be criticized for not answering a question they never sought to answer. What is more, in addition to interpreting quantum mechanics, proponents of $\psi$-*ontic* interpretations could also reconstruct it, for instance based on an operational axiomatization. I agree with all of this but still believe that the success of informational reconstructions puts pressure on $\psi$-*ontic* interpretations as argued by Koberinski & Müller. To see why, consider the following scenario. Assume Peter is a proponent of the many-worlds interpretation. Also assume, he is very interested in quantum reconstruction, realizes that he cannot derive the quantum formalism within the general framework of "theories of many worlds," but successfully derives the quantum formalism from very simple information-theoretic principles. Now assume Jessica is a proponent of some information-based interpretation and that Jessica also successfully reconstructs the formalism from very simple information-theoretic principles. This means that Jessica is in the favorable position that she can both reconstruct and interpret the quantum formalism within a unified conceptual framework. Peter, by contrast, is in the slightly awkward position that he takes a highly mathematical formalism as a given, interprets it in a way that leads to highly counter-intuitive consequences, but derives the formalism from simple information-theoretic principles. Sticking to the dogma of objectivist interpretations that physical theories must never use operational terms, Peter *interprets* vectors in Hilbert space as representing objective reality but *explains* the Hilbert-space structure of the theory in operational terms. In short, assuming that a successful reconstruction is the best way to make sense of the formalism and that successful reconstructions of quantum theory need to be formulated within an operational framework, this forces objectivists to adopt the following thesis: We must formulate and interpret quantum theory in non-operational terms but can make sense of the formalism only within an operational framework.

How can a measurement produce the (apparent) collapse of the wave function? What counts as a measurement? In contrast, the objective of the reconstruction program is to shed light on questions such as: Why does the quantum formalism have the mathematical structure it has? Why is nature so successfully described by the mathematics of complex Hilbert spaces? What are the foundational principles from which the formalism can be derived?

I agree with the reviewer that these are different questions that require different explanatory strategies. For instance, researchers working on reconstructions made much progress by harnessing quantum information theory and identifying and analyzing information-theoretic principles as possible candidates for conceptually foundational principles. Researchers working on interpretations, on the other hand, achieved much progress in specifying different ways in which the wave function can be interpreted, clarifying what the concept of decoherence can and cannot contribute to understanding the collapse, etc. Importantly, however, we must not overlook how intimately the respective questions and programs are connected. In particular, if the quantum reconstruction program succeeds in specifying the information-theoretic principles that constitute the conceptual foundation of quantum theory, it is highly plausible to suppose that this leads to a better "understanding" of quantum theory that is also relevant to the question of what the theory tells us about reality.

Once more, it is helpful to look at special relativity. When we interpret special relativity and ask the question of what it says about the nature of reality, it is perfectly reasonable (at least as a first step) to approach this question by discussing the principles on which the theory is based. This is why I insist that the success of QRP has implications on how to interpret quantum mechanics. As I have argued in this section, the quantum reconstruction program and the endeavor of interpreting quantum mechanics are different but related projects. Of course, one can argue that there are reasons to doubt that QRP will be as successful in providing quantum mechanics with a conceptual foundation as Einstein was regarding special relativity. I address such concerns in the following section. What we note here is that the two projects are intimately connected and that it is thus a virtue if one can approach both projects within a unified conceptual framework.[15]

---

15 A minimal sense in which the success of information-based reconstructions is in tension with objectivist interpretations is that information-based *reconstructions* support information-based *interpretations*. To my knowledge, among the numerous works defending objectivist interpretations, there is not a single one that uses the results of QRP to substantiate their approach. In contrast, one of the very few works explicitly arguing for an "information-theoretic interpretation," (Bub 2018), is driven by the same questions as QRP, namely "What is the fundamental physical principle?" (vi) and "Why the quantum?" (Chapter 9). This exemplifies the close relationship between reconstruction and interpretation, and it should be beyond doubt that successful reconstructions in terms of information-theoretic principles support information-theoretic interpretations. For more details on Bub's information-theoretic interpretation, see in particular Section 10.4 in (Bub 2018) as well as (Dunlap 2022) and (Janas et al. 2022).

## 4. Potential objections to the reconstruction program

It is difficult to defend the reconstruction program because there are so few works critically engaging with it. To give an idea, here is a very incomplete list of recent influential works in philosophy of physics that address the foundations of quantum mechanics but do not even mention the reconstruction program: Adlam 2021, Dürr & Lazarovici 2020, Friebe et al. 2018, Knox & Wilson 2022, Maudlin 2019, Wallace 2021a. By contrast, here is a much more complete list of works that explicitly discuss the reconstruction program: Adlam 2022, Brown & Timpson 2006, Dickson 2015, Felline 2016, French 2023, Grinbaum 2006, 2007, Koberinski & Müller 2018, Letertre 2021. Except for Brown & Timpson 2006 and, with some qualifications, French 2023 and Letertre 2021, these works are highly supportive of the program.

### 4.1 Objection 1: Too many reconstructions

Presently, there are about a dozen of different so-called interpretations of quantum mechanics on the market that have at least some prominent proponents, none enjoying wide agreement, of course. Many view this plethora of interpretations as a symptom of a severe underlying shortcoming. Bohmians, for instance, view this as a shortcoming of (textbook) quantum mechanics, arguing that "the goal of physics must be to formulate theories that are so clear and precise that any form of interpretation […] is superfluous" (Dürr & Lazarovici 2020, viii). Accordingly, Bohmians modify the quantum formalism, believing that the resulting theory provides us with a clear ontological picture. Here the first thing to note is that there is no unique way to modify the quantum formalism. Apart from Bohmian mechanics, the most popular modificatory interpretations are objective collapse theories. So, instead of different interpretations of the same formalism, we have different rival theories – and different interpretations of these rival theories. For instance, Bohmians disagree with each other on the ontological status of the wave function and the space it lives in (Hilbert space or configuration space). Wave function realists come in many different flavors. Some have argued that $3N$-dimensional configuration space is real and that our impression to live in a three-dimensional space is "flatly illusory" (Albert 1996, 277). Others have argued that configuration space is fundamental and three-dimensional space emergent and non-fundamental (Ney 2013). And we also find the idea that there is a fundamental three-dimensional ontology and that the wave function might not be physically real like particles or fields but ontologically sui generis (Maudlin 2013). Of course, this is not a problem just for Bohmian mechanics but for all so-called $\psi$-ontic interpretations (understood in the informal sense introduced in footnote 8). For instance, among proponents of the many-worlds interpretation, we find the most radical endorsement of wave function realism, namely Hilbert space realism (Carroll & Singh 2019), as well as one of the most forceful criticisms of wave function realism, namely Wallace 2021b. This is to say that, at least so far, the promise of getting rid of interpretational problems by replacing the quantum formalism with a rival theory has not been fulfilled.

Of course, proponents of the reconstruction program believe that the plethora of different interpretations is the symptom of a very different shortcoming. This is the shortcoming of the project of interpreting an abstract formalism. Analogously to how the mathematical formalism of special relativity is open to many different interpretations, so is the formalism of quantum mechanics. What special relativity has but quantum mechanics is lacking are clear postulates à la Einstein. However, this invites the following objection: Similar to how we have a number of different interpretations of quantum mechanics, we now have a number of different reconstructions. How exactly is having a number of different reconstructions an improvement over having a number of different interpretations?

My answer to this objection is twofold. First, and most importantly, different reconstructions are not mutually exclusive. This is in stark contrast to interpretations. If Bohmian mechanics is true, the world is governed by deterministic equations. If GRW is true, the world is governed by nondeterministic equations. Either/Or. And either it is true that reality branches in all possible outcomes when a quantum event occurs (some versions of the many-worlds interpretation), or it does not (every non-many-worlds interpretation). The situation is different when it comes to reconstructions. Different reconstructions are not mutually exclusive but rather complementary. Here the mindset is that different reconstructions can shed light on different aspects of quantum theory.[16] Importantly, all successful reconstructions are united by being formulated in an operational framework, specifying information-theoretic principles. As I argued in the previous section, the mere fact that many successful reconstructions can be performed in an operational framework puts pressure on objectivist interpretations.

Secondly, that the same formalism can be derived from different sets of postulates should not be surprising as it is no unique feature of quantum mechanics. Also in special relativity, the role model for the reconstruction program, it is not the case that there exists one single set of meaningful postulates that allows reconstructing the mathematics. As mentioned, Einstein originally derived the mathematical content from the light postulate and the principle of relativity. Alternatively, special relativity can also be based on the single postulate of Minkowski spacetime.[17] In my terminology, deriving special relativity from Einstein's original two principles is the "two-principles reconstruction," deriving it from the postulate of Minkowski spacetime is the "geometrical reconstruction." Corresponding to these reconstructions we have two approaches. An approach, basically, says that its reconstruction is superior in the sense that it can explain the axioms of the other one but not vice versa. My proposal is that both reconstructions are true or legitimate and that both approaches are false or misleading. As in the case of the *quantum* reconstruction program, the idea is that the reconstructions are complementary, shedding light on different aspects or implications of special relativity. The geometrical reconstruction sheds light on the nature of space-time, the two-principles reconstruction is particularly useful when applied to electrodynamics (light postulate) and as a stepping stone to general relativity (from the special principle of relativity to the general principle of relativity).

---

16   See Müller & Masanes 2016, 140. As mentioned in a previous note, Adlam 2022 seems to disagree.
17   And also from the single postulate of universal Lorentz covariance. For a discussion of different approaches, see Brown & Read 2022.

**4.2. Objection 2: Special relativity may be based on physical principles but this does not mean quantum mechanics should be**

One may admit that being based on physical principles is a virtue of special relativity and that this is one of the reasons why special relativity is not plagued by a plethora of diverse interpretations and attempts of modification but insist that this is a unique feature of special relativity and deny that analogous reconstructions can be particularly helpful in elucidating quantum mechanics. Here is how Steven French formulated this worry:

> But the historical development of [quantum mechanics] is more akin to that of General Relativity, with its false starts and application of novel mathematical devices, and to expect to be able to derive the formalism of QM from a few plausible physical postulates might well be regarded as entertaining high hopes. (French 2023, 228)

The first thing to note is that even if it is true that general relativity cannot be reconstructed similarly to special relativity, this does not mean that reconstructing is not a plausible desideratum. We already know that successful reconstructions of quantum theory exist, so why not pay attention to them and attempt to clarify what they imply about quantum mechanics and the world so successfully described by it? Second, analogously to special relativity, (at least) two physical principles played a crucial role in Einstein's discovery of general relativity: the general principle of relativity and the equivalence principle. While it is beyond doubt that these principles played a significant heuristic role in Einstein's thinking, it remains widely contested whether they *should* be considered the foundational principles on which general relativity rests. Einstein insisted they should and many physicists agree. Zeilinger, for instance, states that the equivalence principle plays a similar role as the light postulate in special relativity, saying that "[b]oth foundational principles are very simple and intuitively clear" (Zeilinger 1999, 632). Others have claimed that Einstein's version of the equivalence principle is even inconsistent with general relativity (see Lehmkuhl 2022; Norton 1993).[18] It would go beyond the scope of this paper to engage in this discussion. Instead, I stress that while many commentators argue that Einstein's original principles are problematic, various alternative formulations have been formulated that could get the job done. These alternative formulations are often also based on an underlying physical principle such as Anderson's "principle of general invariance," discussed in Norton 1993.

I conclude this section by addressing classical mechanics. Critics of the quantum reconstruction program may argue that classical mechanics does not seem to be in need of being reconstructed. This is true in a sense. But the reason why we do not need to reconstruct classical mechanics is that classical mechanics developed in a way such that it has been shaped by underlying physical principles from the very beginning (see Goyal 2023 and Darrigol 2014).

As pointed out above, it is important to note that classical mechanics can be formulated in different and highly abstract ways. Typically, if we have a system of N particles, we say that the "state of the system is represented by $N$ points $X_1,...X_N$, in three-dimensional Euclidean space" (Wallace 2021b, 68). But this is only one possible mathematical representation. As Wallace points out "there is another way to represent this theory. We can define the configuration space as the product of $N$ copies of Euclidean three-space. Each $N$-tuple of points $(X_1,...X_N)$ now corresponds to a single point in this $3N$-dimensional space" (Wallace 2021b, 68). Wallace brings this example as an argument against wave function realism, highlighting that it would be misleading to consider this $3N$-dimensional configuration space as the true space we live in. I agree. I would add that this shows how problematic it can be to take the mathematical representation of some theory at face value, trying to read off an ontology directly. In the present context, I only want to highlight that if, for whatever reason, classical mechanics had developed such that first we had formulated the theory in $3N$-dimensional configuration space and only later realized it can be reformulated in three-dimensional space, this reconstruction would have been very useful to make sense of the physical meaning of the theory. It is also noteworthy that Newton's three laws of motion certainly qualify as

---

18 "Einstein was adamant in defending his version of the principle against much of the rest of the community, and against the claim that it was of *only* heuristic importance in the search for GR. As we shall see, for Einstein it was also intimately related to what he saw as the main result of GR: the unification of inertia and gravity in a sense to be specified. In contrast, others argued that the Einstein equivalence principle is *false* according to GR, but that one version or the other of the strong equivalence principle (related to the local validity of special relativity) *does* hold in the theory." (Lehmkuhl 2022, 126) physical principles but they are not explicit in the reformulations of Lagrangian or Hamiltonian mechanics. Certainly, if one were only familiar with Lagrangian mechanics, one would have a very different picture of classical mechanics (and it would be misleading to consider the Lagrangian a physically real quantity).

**4.3. Objection 3: Intuitive physical principles?**

The quest of QRP to derive the quantum formalism from physical principles is occasionally stated as looking for *intuitive* physical principles. Zeilinger, as mentioned above, calls the principles underlying special and general relativity "very simple and intuitively clear" as well as "intuitively nearly obvious principles" (Zeilinger 1999, 632). This may invite the following objection:

> "It is not always clear what it is that the advocates of reconstructive approaches are seeking for their physical basis – it cannot be something 'intuitive' because that way lies denial of the impact of modern physics on our intuitions and, again appealing to the analogy with relativity theory, the principle of the constancy of the speed of light is certainly not intuitive. And relatedly, such approaches should abandon the effort to build on the back of classical mechanics, as we have just noted, in order to bring them closer to the practice of physics."[19]

I agree that the terminology of "intuitive" principles is misleading. It is thus crucial to point out that when proponents of QRP use this terminology, they usually do not want to say that the principles they are looking for are intuitive in the sense of immediately compelling. Instead, they are supposed to be "intuitively graspable" (Goyal 2023) in the sense that it is easily graspable what they mean. The statements "A moving body's measured length is shorter than the length measured in the body's own rest frame" and "Whether two spatially separate events take place simultaneously cannot be decided absolutely but depends on the observer's reference frame" may be highly counter-intuitive but are still intuitively graspable. A person who is unfamiliar with special relativity and whose physical intuitions are strongly shaped by our everyday encounter with the world will find these statements hard to believe and in tension with her experience. Still, she will be able to get some understanding of what they mean since they have meaningful contents that can be expressed in fairly "normal" non-technical language. Given that quantum mechanics is notorious for having a number of counter-intuitive consequences it would indeed be surprising if it could be derived from a set of prima facie compelling principles. This is precisely why it is often considered a desideratum to find an underlying principle that is "surprising" or "paradoxical" (Koberinski & Müller 2018, 278).

---

19  This passage is from an early 2023 draft version of Steven French's new book *A Phenomenological Approach to Quantum Mechanics* (2023). This passage was later cut.

## 4.4. Objection 4: Quantum reconstructions are only principle theories

In 1919 Einstein published a popular article entitled "What is the Theory of Relativity?" in *The Times* (London). Here he made the famous distinction between principle theories and constructive theories. A principle theory comprises a set of empirically well-confirmed generalizations/principles, accounting for certain physical phenomena in the sense that these phenomena follow from the respective principles. According to Einstein, prime examples are thermodynamics and his theory of relativity. A constructive theory can account for a wide range of diverse phenomena by reducing them to the same underlying mechanism. Einstein's prime example is the kinetic theory of gases (or, perhaps more appropriately, statistical mechanics). Proponents of QRP occasionally argue that quantum reconstructions are to be viewed as principle theories (Clifton et al. 2003, Koberinski & Müller 2018). Now, it has been argued that if quantum reconstructions are principle theories, this is bad news for the foundational ambitions of QRP (Brown & Timpson 2006). This is because constructive theories are more fundamental than principle theories. If you are interested in the nature of temperature, for instance, you should consult statistical mechanics, not thermodynamics. It has been noted that Einstein himself seemed to question whether principle theories can have explanatory power and regarded them as stepping stones toward constructive theories (Brown & Timpson 2006). This is why Brown and Timpson argue that it is unwise to consult special relativity as a template "for a *fundamental* interpretation."

Here are four reasons to resist the objection raised by Brown & Timpson. The first two points are modest and suffice to show that QRP is at least a worthwhile project. 3. and 4. point toward more radical approaches.

1. It is contested whether Einstein really believed that principle theories lack explanatory power (Lange 2014). More importantly, there is good evidence that he believed that special relativity has explanatory power (Lange 2014). Most importantly, regardless of what Einstein believed, it is certainly reasonable to assume that his two principles play an important role in explaining the phenomena of special relativity.

2. Subsequently, if quantum reconstructions can do for quantum mechanics what Einstein did for special relativity, QRP can be considered a great success.

3. Einstein may have expected that special relativity becomes replaced by a constructive theory but this constructive theory never showed. The recent success in reconstructing quantum theory can be understood as suggesting that principle theories are foundationally more significant than expected by Einstein (Grinbaum 2006; see also Adlam 2022, Section 5.3).

4. One may question whether the distinction between principle theories and constructive theories is particularly useful. More interestingly, perhaps, one may question whether quantum reconstructions should be understood as principle theories in Einstein's sense. This is because, as noted by Grinbaum, Einstein consistently believed that any kind of physical theory "should describe 'the real state of the real system'" (Grinbaum 2017). However, as we noted in Section 3.3, quantum reconstructions put pressure on objectivist and *ψ-ontic* interpretations. Following Grinbaum 2017, one might argue that quantum reconstructions are not stepping stones toward constructive theories but rather stepping stones toward theories that "do away with the idea of entities" (see also Adlam 2022).[20]

### 4.5. Objection 5: The success of information-based reconstructions is no surprise because we have arrived at quantum mechanics on the basis of information

An anonymous reviewer of this journal argued that my thesis defended in Section 3.3, namely that the success of informational reconstructions puts pressure on objectivist interpretations, can only be plausible if the success of such reconstructions is somehow surprising. Here is one of the reviewer's arguments of why they think this success is no surprise.

> "[I]t is not surprising that informational concepts come in useful in axiomatizing quantum mechanics, because we have arrived at quantum mechanics on the basis of information (i.e. empirical data) and so it is natural that we can come up with characterisations of the theory which are informational – they are informational because they are operational and thus they are about *the empirical data*, not because quantum mechanics itself is necessarily 'about' information."

I would summarize this objection as follows: When we construct a scientific theory, we typically do so in the face of information we gain by experimenting on the physical world. Thus, it should not come as a surprise that we can reconstruct the respective theory in terms of information. This objection is potentially very powerful because it rests on an intuitive and universally accepted thesis of how we typically arrive at scientific theories. However, I believe that this objection can be conclusively rebutted. What is more, I believe that the way this objection should be rebutted allows me to further strengthen my case that informational reconstructions put pressure on objectivist interpretations.

---

20  This seems to cohere nicely with Mittelstaedt's claim that the success story of modern physics can be constructed as a story of abandoning metaphysical hypotheses (Mittelstaedt 2011).

The first and most important thing to note in this context is that reconstructions of other physical theories exist, e.g., of classical mechanics, and that these reconstructions are *not* based on information-theoretic axioms. Reconstructions of classical mechanics are discussed, e.g., in Goyal 2020, 2023 and Darrigol 2014, 2020. Goyal's reconstruction is formulated within an "energetic framework" such that the theory is derived from a principle of conservation and the Galilean principle of relativity (Goyal 2020). As pointed out by Goyal, the Newtonian framework rests on the idea that there are quantities that are in principle unobservable, such as absolute space and time, and that an ideal observer could passively register all relevant properties (Goyal 2023). This is in stark contrast to the operational character of quantum mechanics and this, in turn, is reflected in the successful information-based reconstructions discussed above.

Accordingly, I respond to the present objection as follows: Classical mechanics and quantum mechanics are both constructed on the basis of information we obtained by experimenting on the physical world. But only in the case of quantum mechanics informational reconstructions proved useful in elucidating the formalism. This implies that the fact that we have arrived at quantum mechanics on the basis of information does not mean that the success of informational reconstructions is trivial. On the contrary, it seems to be a distinctive feature of quantum mechanics that information-theoretic principles can play such a foundational role. This brings us back to the question of why QRP is so successful and the answer suggested in Section 3.3, namely that quantum mechanics, fundamentally, is a theory about information. This case can be strengthened by the following observation. While there exist several successful reconstructions that are based on information-theoretic principles, none exists that is based on the framework of objectivist interpretations such as Bohmian mechanics or the many-worlds interpretation. Here is how Fuchs made this point:

> "[T]he idea of quantum states as information has a simple unifying power that goes some way toward explaining why the theory has the very mathematical structure it does. By contrast, who could take the many-worlds idea and derive any structure of quantum theory out of it? This would be a bit like trying to regrow a lizard from the tip of its chopped-off tail: The Everettian conception of never purported to be more than a reaction to the formalism in the first place." (Fuchs 2014, 388)

Of course, as anticipated in this quote and discussed in footnote 12, proponents of objectivist interpretations can emphasize that they never understood their interpretation as being capable (or in need) of explaining why quantum theory has the distinctive mathematical structure it has. However, if objectivist interpretations operate within frameworks that do not prove useful in the context of reconstructing quantum theory, while, by contrast, non-objectivist information-based interpretations operate within a framework that proves useful, this is a virtue of non-objectivist interpretations and thus the success of informational reconstructions puts pressure on objectivist interpretations.

We can further emphasize the usefulness of *information*-based reconstructions by considering the history of QRP. As discussed in (Darrigol 2015, Grinbaum 2006, 2007, and D'Ariano et al. 2017, 2-4), attempts to reformulate quantum theory in a way such that the abstract Hilbert-space structure can be understood as a consequence of simpler principles have existed since the early days of quantum axiomatization. Von Neumann himself tried to do so in the joint work with Birkhoff on quantum logic. This work has been taken up, e.g., by Paulette Destouches-Février, Constantin Piron, Josef Jauch, and George Mackey. Especially interesting are the reconstructive works by Günther Ludwig that culminated in Ludwig 1985. Both Ludwig and Mackey "shifted the foundational basis from quantum logic to the structure of a probabilistic state space" (Darrigol 2015, 329), but Ludwig in particular aimed at an operational approach and his axiomatization is noticeably closer to the modern informational reconstructions. Importantly, however, all these early axiomatizations "made little impact" (Goyal 2023, 369) and "more insightful axiomatization[s] re-emerged with the rise of quantum information" (D'Ariano et al. 2017, 3), Hardy's information-based reconstruction (2001) constituting the "real turning point" (Darrigol 2015, 329). This brings me to a further worry of the reviewer as to why the success of informational reconstructions should not come as a surprise:

> "Also, while it is true that a lot of reconstructions of QM are informational, there seems to be some reason to think that this is a sociological fact rather than a reflection of something deep about the theory – quantum information is currently in vogue, and also many people working on QRP come from the quantum information community so they are working in the language they are familiar with. It's not clear to me that we should draw stronger conclusions than this."

The reviewer is certainly right that working on quantum information is currently in vogue and that this is among the reasons why there exist several reconstructions based on this concept. However, as pointed out above, reconstructive attempts have existed since the 1930s. Given the development of QRP, it is safe to say that informational reconstructions turned out to be particularly useful. Accordingly, we can summarize the relationship between quantum theory and informational reconstructions as follows:

1. It turns out that quantum theory in particular is well-suited to be reconstructed in terms of information-theoretic principles (as opposed to, e.g., classical mechanics).

2. It turns out that the framework of information theory proves particularly useful for reconstructing quantum theory (as opposed to, e.g., the framework of quantum logic or the abstract reconstructions of Mackey and Ludwig).

Again, of course, this does not imply that quantum theory is, fundamentally, a theory about information but it suggests that we should consider this a more serious alternative than it is typically done in philosophy of quantum mechanics.

## 4.6. Objection 6: The success of QRP is no surprise because quantum mechanics can be defined in operational terms

Another reason why one of the reviewers of this journal argued that the success of QRP is no surprise is the fact that quantum mechanics is a theory that can be written in operational terms. Here is the argument:

> "In QM we have a theory which can be defined in operational terms and which exhibits certain observable regularities. Surely it is to be expected that there would be some possible ways of systematizing those observable regularities in terms of 'principles' – almost by definition, regularities can always be systematized in some way. Indeed I would be much more surprised to find that some scientific theory which can be written in operational terms does not have an operational axiomatization! […] [T]he counter-intuitive consequences [of quantum mechanics] are mostly noticeable once one tries to give an ontic account of it –  e.g. one has to say something about ontic structure if one wants to say anything about contextuality, nonlocality, wave-particle duality. It doesn't seem to me surprising that if one refrains from talking about ontic structure and just focuses on the operational consequences, then there are simple and reasonably intuitive ways of systematizing these purely operational facts."

This brings us back to Section 3.3 where I pointed out that one of the characteristic features of (textbook) quantum mechanics is that an operational term such as "measurement" plays a central role and that this does not sit well with many philosophers of physics who typically favor an objectivist interpretation in which the formalism is purged of all operational notions. I also noted that purging quantum mechanics from the notion of measurement comes at a cost. This cost manifests in that one either has to accept many worlds or modify the quantum formalism. As argued above, a further cost is that by subscribing to an objectivist interpretation you move from an operational framework that allows for successful reconstructions to a framework that does not offer this kind of explanatory power. For instance, as mentioned in the previous subsection, no reconstruction exists that operates in the framework of many worlds. Accordingly, when the reviewer says that "the counter-intuitive consequences [of quantum mechanics] are mostly noticeable once one tries to give an ontic account of it" I consider this as supporting my claim that we should more seriously entertain the idea that quantum mechanics is not primarily about representing external reality (as it is the case for *ψ-ontic* interpretations). Instead, as suggested in this paper, we may understand quantum mechanics to be fundamentally about information. In this context, it is to be noted that non-objectivist interpretations such as Bohr's version of the Copenhagen interpretation, QBism, or Healey's pragmatism are not committed to non-local dynamics, i.e. spooky action at a distance, while objectivist interpretations such as Bohmian mechanics and GRW imply this counter-intuitive feature.[21]

**4.7. Objection 7: The notion of information**

In this paper, I have emphasized repeatedly that QRP has received unreasonably little attention in the philosophy community. This is true for quantum information theory in general. Among the notable exceptions is Christopher Timpson's seminal monograph *Quantum Information Theory and the Foundations of Quantum Mechanics* (2013). Here Timpson critically discusses how the concept of information is used in quantum information theory. One of his conclusions is that "informational immaterialism," the view that information is ontologically fundamental, is untenable. I want to stress that QRP does not imply such a view and I doubt that proponents of QRP subscribe to it (Zeilinger might be an exception). Timpson also argues that often the technical sense and the everyday sense of the concept of information are conflated which might lead to confusion. One might argue that, in further consequence, the concept of information is problematic and that thus information-theoretic reconstructions are conceptually flawed. However, Timpson argues that a clear-cut distinction between the technical and the everyday sense is possible and offers an analysis of each term. Accordingly, I don't think that Timpson's approach to the concept of (quantum) information can be used to launch such an attack against QRP. There is a further, more general, objection related to the concept of information: One might believe that informational reconstructions lead to *ψ-epistemic* interpretations and are thus refuted by the PBR theorem. This will be addressed in the following subsection.

---

21 The case is less clear in RQM and MWI. Proponents of RQM have argued that by abandoning "strict Einstein realism" RQM can reconcile quantum mechanics with "completeness, (operationally defined) separability, and locality" (Smerlak & Rovelli 2007, 427). However, RQM might still be too objectivist to avoid non-locality (see Pienaar 2019). In the case of MWI, there is some consensus that MWI avoids action at a distance (see, e.g., Vaidman 2021). However, if "global branching" is true as assumed by Sebens and Carroll, then MWI "implies that observers here on Earth could be (and almost surely are) branching all the time, without noticing it, due to quantum evolution of systems in the Andromeda Galaxy and elsewhere throughout the universe" (Sebens & Carroll 2018, 35).

**4.8. Objection 8: Is QRP in tension with the PBR theorem?**

Let us assume the results in Section 3.3 are correct and QRP puts considerable pressure on *ψ-ontic* interpretations. Where does this leave us? It is often assumed that the only alternative to *ψ-ontic* interpretations are *ψ-epistemic* interpretations. (Here I understand *ψ-epistemic* interpretations in the informal sense of interpreting the wave function as representing one's knowledge of the respective physical system.) And indeed this is the path suggested by Koberinski & Müller 2018. However, it is widely acknowledged that the PBR theorem (Pusey et al. 2012) rules out *ψ-epistemic* models in the strict sense of (Haarigan & Spekkens 2010). Furthermore, *ψ-epistemic* interpretations have recently been challenged also on conceptual grounds (Luc 2023). Does this mean that QRP leads us to a dead end? In what follows, I specify two ways out. Option 1 is for proponents of QRP who believe that QRP strongly discourages *ψ-ontic* interpretations. It exploits the fact that *ψ-ontic* and *ψ-epistemic* interpretations are not the only options. Option 2 is for proponents of QRP who believe that QRP strongly encourages *ψ-epistemic* interpretations. It exploits the fact that *ψ-ontic* and *ψ-epistemic* interpretations are not mutually exclusive options (if understood in the informal sense).

Option 1: An alternative to *ψ-ontic* and *ψ-epistemic* interpretations are what I call *ψ-doxastic* interpretations. A *ψ-doxastic* interpretation says that the wave function neither represents an ontic state nor the knowledge about the underlying ontic state but the agent's degrees of belief. The most prominent version of a *ψ-doxastic* interpretation is QBism. The distinctive idea of QBism is to apply a personalist Bayesian account of probability to quantum probabilities (Fuchs et al. 2014). Accordingly, QBism argues that quantum states do "not represent an element of physical reality but an agent's personal probability assignments, reflecting his subjective degrees of belief about the future content of his experience" (Fuchs & Schack 2015, 1). We see that QBists clearly reject the ideas that the wave function should be reified or objectified and that the wave function represents objective reality. This is why it is often believed that QBism is a *ψ-epistemic* interpretation (see, e.g., Koberinski & Müller 2018). But this is misleading. The QBist claim that the wave function represents degrees of belief amounts to a *doxastic* and not an *epistemic* interpretation of the quantum state/wave function. This is because knowledge is a factive notion. If one knows that *p*, then *p* is the case. Importantly, this is *precisely* why QBism avoids the PBR no-go theorem. While the PBR theorem rules out *ψ-epistemic* interpretations, it is silent on the QBist claim that wave functions represent degrees of beliefs about one's future experiences (DeBrota & Stacey 2019, Glick 2021, Hance et al. 2022). Unfortunately, in the literature the PBR theorem is often misunderstood as ruling out any interpretation that is not *ψ-ontic* (e.g. Maudlin 2019, 83-89). This means overlooking how the QBist escapes the PBR theorem. Of course, going down this road means to radically break with how we are used to understand scientific theories. But it constitutes the most consistent way for avoiding *ψ-ontic* interpretations and it is often assumed that QBism is the best-developed interpretation in the Copenhagen spirit.

> "In many ways, quantum Bayesianism represents the *acme* of certain traditional ways of thinking about quantum mechanics (broadly speaking, Copenhagen-inspired ways). If one hopes to defuse the conceptual troubles over collapse and nonlocality by conceiving of the quantum state in terms of some cognitive state, then the only satisfactory way to do so is by adopting the quantum Bayesian line." (Timpson 2013, 7f.)

Option 2: Hance, Rarity, and Ladyman have recently pointed out that "there is no reason to suppose that a one-one map between the wavefunction and the ontic state rules out that the wavefunction represents knowledge" (2022). This is to say that interpretations can be both *ψ-ontic* and *ψ-epistemic* (although not in the strict sense introduced in Harrigan and Spekkens 2010). One approach to quantum mechanics that exploits this fact is Steven French's phenomenological interpretation based on the work of Fritz London and Edmond Bauer (French 2023, Section 10.5). This option promises the best of both worlds. To account for the success of reconstructing quantum theory in terms of information and knowledge, and to be in accordance with how we typically interpret scientific theories. Accordingly, philosophers with strong sympathies for *ψ-ontic* interpretations need not worry that QRP is inconsistent with their interpretative endeavor. Of course, option 2 is widely uncharted territory. A possible objection would be that choosing this option means watering down what might be the most interesting implications of QRP.

**Conclusion**

The quantum reconstruction program (QRP) has been unreasonably ignored in contemporary philosophy of physics. This is unfortunate because successful reconstructions can help philosophers in their quest for interpreting quantum mechanics and reconstructions are themselves in need of philosophical reflection and interpretation. In this paper, I discussed philosophical implications of QRP and defended it against possible objections. I argued that successful reconstructions put pressure on the most prominent interpretations we find in philosophy, i.e., so-called "quantum theories without observers" (Section 3.3). In Section 4, I addressed eight objections against QRP. In the subsections 4.4 and 4.8, I sketched two possible ways in which QRP could develop. According to the more moderate one, quantum reconstructions can be understood as principle theory's in Einstein's sense and are consistent with $\psi$-ontic interpretations. According to the more radical one, QRP might want to break with the traditional assumption that physical theories describe "the real state of the real system" and pave the way for $\psi$-doxastic interpretations.

**Acknowledgments:** A previous version of this paper was discussed in a reading group session at the Center for Philosophy of Science at the University of Pittsburgh. I'm particularly grateful for the feedback from Adam Koberinski, Eleanor Knox, Lucy Mason, John Norton, and David Wallace. I'm most indebted to Philip Goyal for convincing me of the significance of the quantum reconstruction program and for many helpful discussions and clarifications. This research was funded in part, by the Austrian Science Fund (FWF) [10.55776/P36542].